\documentstyle[12pt]{article}

\font\twelve=cmbx10 at 15pt
\font\ten=cmbx10 at 12pt
\font\eight=cmr8

\def\upa{\uparrow}
\def\downa{\downarrow}
\def\cit#1{$^{[#1]}$}
\def\pr{Phys.\ Rev.\ }
\def\prl{Phys.\ Rev.\ Lett.\ }
\def\pl{Phys.\ Lett.\ }
\def\etal{{\it et al}.}

\begin{document}

\begin{titlepage}

\begin{center}

{\ten Centre de Physique Th\'eorique\footnote[2]{
Unit\'e Propre de Recherche 7061} - CNRS - Luminy, Case 907}
{\ten F-13288 Marseille Cedex 9 - France }

\vspace{1 cm}

{\twelve FERMI-DIRAC DISTRIBUTIONS}
{\twelve FOR QUARK PARTONS}

\vspace{0.3 cm}

{\bf C. BOURRELY, F. BUCCELLA\footnote[1]{Dipartimento di Scienze
Fisiche, Universit\'a di Napoli, Mostra d'Oltremare, Pad. 19, 80125
Napoli, Italy}, G. MIELE$^{\hbox{\eight *}}$, \\
G. MIGLIORE$^{\hbox{\eight *}}$, J. SOFFER and V.
TIBULLO$^{\hbox{\eight *}}$}

\vspace{1 cm}

{\bf Abstract}

\end{center}

We propose to use Fermi-Dirac distributions for quark and antiquark
partons. It allows a fair
description of the $x$-dependence of the very recent NMC data on the
proton and neutron structure
functions $F_2^p(x)$ and $F_2^n(x)$ at $Q^2=4$ GeV$^2$, as well as
the CCFR antiquark distribution
$x\overline q(x)$. We show that one can also use a corresponding
Bose-Einstein expression to describe
consistently the gluon distribution. The Pauli exclusion principle,
which has been identified to
explain the flavor asymmetry of the light-quark sea of the proton,
is advocated to guide us for
making a simple construction of the polarized parton distributions.
We predict the spin dependent
structure functions $g_1^p(x)$ and $g_1^n(x)$ in good agreement with
EMC and SLAC data. The quark
distributions involve some parameters whose values support well the
hypothesis that the violation of
the quark parton model sum rules is a consequence of the Pauli
principle.

\vspace{1 cm}

\noindent Number of figures : 6

\bigskip

\noindent October 1993

\noindent CPT-93/P.2961

\bigskip

\noindent anonymous ftp or gopher: cpt.univ-mrs.fr

\end{titlepage}

\section{Introduction}

Many years ago Feynman and Field made the conjecture\cit{1} that the
quark  sea in the proton may not be flavor symmetric, more precisely
$\bar d>\bar u$, as a consequence of Pauli principle which favors $d\bar
d$ pairs with respect to $u\bar u$ pairs because of the presence of two
valence $u$ quarks and only one valence $d$ quark in the proton. This
idea was confirmed by the results of the NMC experiment\cit{2} on the
measurement of proton and neutron  unpolarized structure functions,
$F_2(x)$. It yields a fair evidence for a defect in the Gottfried sum
rule\cit{3} and one finds

\begin{equation}
I_G = \int^1_0 \frac{dx}{x} [ F^p_2(x)-F^n_2(x)] = 0.240\pm0.016
\label{1}
\end{equation}

\noindent instead of the value 1/3 predicted with a flavor symmetric
sea, since we have in fact

\begin{equation}
I_G = \frac{1}{3} (u+\bar u - d -\bar d) = \frac{1}{3} + \frac{2}{3}
(\bar u - \bar d). \label{2}
\end{equation}

\noindent A crucial role of Pauli principle may also be advocated to
explain the well known dominance of $u$ over $d$ quarks at high
$x$,\cit{4} which explains the rapid decrease of the ratio $F^n_2(x)
/ F^p_2(x)$ in this region. Let us denote by $q^\upa (q^\downa)$, $u$ or
$d$ quarks with helicity parallel (antiparallel) to the proton helicity.
The double helicity asymmetry measured in polarized muon (electron) -
polarized proton deep inelastic scattering allows the determination of
the quantity $A^p_1(x)$ which increases towards one for high $x$,\cit{5}
suggesting that in this region $u^\upa$ dominates over $u^\downa$, {\it
a fortiori} dominates over $d^\upa$ and $d^\downa$, and we will see now,
how it is possible to make these considerations more quantitative.
Indeed at $Q^2=0$ the first moments of the valence quarks are related to
the values of the axial couplings

\begin{equation}
u^\upa_{\rm val} = 1+F, \quad u^\downa_{\rm val}=1-F, \quad d^\upa_{\rm
val} = \frac{1+F-D}{2}, \quad d^\downa_{\rm val} = \frac{1-F+D}{2},
\label{3}
\end{equation}

\noindent so by taking $F=1/2$ and $D=3/4$ (very near to the quoted
values\cit{6} $0.477\pm0.011$ and $0.755\pm0.011$) one has $u^\upa_{\rm
val} =3/2$ and $u^\downa_{\rm val} = 1/2$ which is at the center of the
rather narrow range $(d^\upa_{\rm val},d^\downa_{\rm val}) = (3/8,
5/8)$. The abundance of each of these four valence quark species,
denoted by $p_{\rm val}$, is given by eq.~(\ref{3}) and we assume that
the distributions at high $Q^2$ ``keep a memory'' of the properties of
the valence quarks, which is reasonable since for $x>.2$ the sea is very
small. So we may write for the parton distributions

\begin{equation}
p(x) = F(x, p_{\rm val}) \label{4} \end{equation}

\noindent where $F$ is an increasing function of $p_{\rm val}$. The fact
that the dominant distribution at high $x$ is just the one corresponding
to the highest value of $p_{\rm val}$, gives the correlation
{\it abundance -
shape} suggested by Pauli principle, so we expect broader shapes for
more abundant partons. If $F(x, p_{\rm val})$ is a smooth function of
$p_{\rm val}$, its value at the center of a narrow range is given, to a
good approximation, by half the sum of the values at the extrema, which
then implies\cit{7}

\begin{equation}
u^\downa(x) = \frac{1}{2} d(x) . \label{5} \end{equation}

\noindent This leads to

\begin{equation}
\Delta u(x) \equiv u^\upa(x) - u^\downa(x) = u(x) - d(x)\label{6}
\end{equation}

\noindent which allows to relate the contribution of $u$ quark to
$g^p_1(x)$, the proton polarized structure function, to the
contributions of $u$ and $d$ quarks to $F^p_2(x) - F^n_2(x)$ i.e.

\begin{equation}
xg^p_1(x) \biggl|_{u} = \frac{2}{3} \left(F^p_2(x) - F^n_2(x)
\right) \biggr|_{u+d}. \label{7} \end{equation}

\noindent Moreover, since $\Delta d_v = (\frac{F-D}{2F}) \Delta u_v
= -1/4 \Delta u_v$ and $e^2_d=1/4 e^2_u$, we expect a small negative
correction
to $g^p_1(x)$ from $d$ quarks and therefore

\begin{equation}
xg^p_1(x) \simeq \frac{2}{3} \left( F^p_2(x) - F^n_2(x) \right),
\label{8} \end{equation}

\noindent at least in the $x$  region dominated by valence quarks. This
is confirmed by experiment\cit{2,5}, which shows good agreement with
eq.~(\ref{8}) for $x\ge.2$, with a slight overestimate of $g^p_1(x)$ at
small $x$, partly due to the fact that we have neglected the $d$ quark
contribution $\Delta d$, which is expected to be negative \cit{7}. All
these features are encouraging to take rather seriously the importance
of Pauli principle and have led us to propose a new description of the
quark parton distributions in terms of Fermi-Dirac expressions as
functions of the scaling variable $x$ according to\cit{8}

\begin{equation}
p(x)  = \frac{f(x)}{e\frac{x-\tilde x(p)}{\bar x} + 1}\qquad. \label{9}
\end{equation}

\noindent Here $\tilde x(p)$ plays the role of the ``thermodynamical
potential'' for the fermionic parton $p$ of a given flavor and helicity,
$\bar x$ is the ``temperature'' and $f(x)$ is a weight function whose
support is the open range $(0,1)$. As we will see, we will be able to
reproduce the NMC data\cit{9} for $F^p_2(x)$ and $F^n_2(x)$ taken at
$Q^2=4$ GeV$^2$, and the antiquark data from neutrino deep inelastic
scattering \cit{10}.
This will
allow the determination of the small number of parameters entering in
eq.~(\ref{9}). Keeping the same
value for
$\bar x$, we will also produce a gluon distribution $G(x)$ in terms of a
Bose-Einstein expression similar to eq.~(\ref{9}) and consistent with
what is known from experiment\cit{11,12}. We will find for $\tilde
x(u^\upa)$ a much larger value than for $\tilde x(u^\downa)$ (i.e.\
$\tilde x(u^\upa)\sim 4\tilde x(u^\downa)$) which suggests that the
defect in the Gottfried sum rule is a consequence of the fact that the
``$u^\upa$ bus'' is almost full and therefore the sea quarks have to
fill up all available space in the ``$u^\downa$ bus'' and in the ``$d$
bus''. This allows us to propose an interesting solution of the spin
crisis, where we will have a fair description of $g^p_1(x)$ obtained by
EMC\cit{5} and also of the neutron polarized structure function
$g^n_1(x)$ obtained very recently at SLAC\cit{13}.

In the next section we present our new approach for parton distributions
and we will specify our parameterization for $u$ and $d$ quarks, for
antiquarks and for gluons. In section~3, by using the most recent and
accurate available low $Q^2$ unpolarized deep inelastic data, we will
see that it is possible to complete the determination of all these
distributions in terms of eight free parameters and we will also present
our predictions for polarized deep inelastic scattering. In section~4,
we will discuss shortly the outcoming results of this approach for
 some quark parton sum rules. We will give our
concluding remarks in section~5.

\section{New Approach to Parton Distributions}

Let us first consider $u$ quarks and antiquarks and we assume that
$u^\upa(x)$, $u^\downa(x)$, $\bar u^\upa(x)$ and $\bar u^\downa(x)$ are
expressed in terms of Fermi-Dirac distributions of the form given by
eq.~(\ref{9}). The thermodynamical potential $\tilde x(p)$ is a constant
parameter which is expected to be different for each parton $p$, whereas
the temperature $\bar x$ can be a universal constant and we will take
for $f(x)$ a simple form\footnote{With this choice $p(x)$ does not go
to zero when $x\to
1$, but it has a fast decrease coming from the exponential in the
denominator.}

\begin{equation}
f(x) = A x^\alpha. \label{10} \end{equation}

\noindent Concerning the $d$ quarks and antiquarks, following our above
arguments we will assume that

$$ d(x) = \frac{u^\downa(x)}{1-F} \eqno(5^\prime) $$

\noindent which is a slight modification of eq.~(\ref{5}), to account
for the fact that $F$ is not exactly 1/2. As also indicated above,
$\Delta d(x)$ is not very large and moreover the potentials associated
to $d^\upa(x)$ and $d^\downa(x)$ are expected to satisfy the following
constraints

\begin{equation}
0<\tilde x(d^\downa)-\tilde x(u^\downa) \simeq \tilde x(u^\downa) -
\tilde x(d^\upa) < \bar x. \label{11} \end{equation}

\noindent Therefore this observation justifies, to a reasonable
approximation, the following choice

\begin{equation}
\Delta d (x) = -k f(x) \frac{e\frac{x-\tilde x(u^\downa)}{\bar
x}}{\left( e\frac{x-\tilde x(u^\downa)}{\bar x} +1\right)^2}, \label{12}
\end{equation}

\noindent an expression with no new parameter, except the normalization
factor $k$ which will be fixed by requiring for the first moment

\begin{equation}
\Delta d = \Delta d_{\rm val} = F-D. \label{13}
\end{equation}

\noindent We therefore assume that the $d$ sea quarks are not
polarized\footnote{Clearly whereas one
needs a large negative polarization for $u$ sea quarks and antiquarks,
it is less crucial for the
case of the $d'$s which where assumed to be slightly polarized in
refs.[14] and [15].} and
consistently we also take for the
$d$ antiquarks,

\begin{equation}
\bar d^\upa(x) = \bar d^\downa(x) \equiv \bar u^\downa(x). \label{14}
\end{equation}

\noindent Concerning the strange quarks we first take in accordance with
the data\cit{10}

\begin{equation}
s(x) = \bar s(x) = \frac{\bar u(x) + \bar d(x)}{4}. \label{15}
\end{equation}

\noindent Finally for the gluon distribution, for the sake of
consistency, one should assume a Bose-Einstein expression given as

\begin{equation}
G(x) = \frac{16}{3}  \frac{f(x)}{e\frac{x-\tilde x(G)}{\bar x} -1}
\label{16} \end{equation}

\noindent with the same temperature $\bar x$, a specific potential
$\tilde x(G)$ and where the factor 16/3 is just twice the ratio of the
color degeneracies of gluon and quarks since $G(x)$ is the unpolarized
gluon distribution.

This completes the parameterization of all parton distributions we will
use and we will now proceed to the determination of the eight parameters
we have introduced by means of the description of low $Q^2$ unpolarized
deep inelastic scattering data. We will also give the predictions it
leads to for the polarized structure functions $g^p_1(x)$ and
$g^n_1(x)$.

\bigskip

\section{Unpolarized and Polarized Deep Inelastic Scattering}
\medskip

To determine our parameters we have used the most recent NMC
data\cit{9}\footnote{In ref.[9] the
result quoted in eq.(1) has been slightly changed to $I_G=0.258
\pm0.017$.} on $F^p_2(x)$ and
$F^n_2(x)$ at
$Q^2=4$ GeV$^2$, together with the most accurate CCFR data on the
antiquark distribution [10].

\noindent We get the following set of parameters

\begin{equation}
\bar x=0.132\ ,\ A=0.579\ ,\ \alpha=-0.845,\label{17}
\end{equation}
$$\tilde x(u^{\upa})=0.524\ ,\ \tilde x(u^{\downa})=0.143\ ,\
\tilde x(\bar u^{\upa})=-0.216\ ,\
\tilde x(\bar u^{\downa})=-0.141.$$
For the fraction of the total momentum carried by quarks and antiquarks
we find
\newpage
$$\int^{1}_{0} xu(x)dx=0.278\ ,\ \int^{1}_{0} xd(x)dx=0.075,$$
\begin{equation}
\int^{1}_{0} x[\bar u(x)+\bar d(x)+s(x)+\bar s(x)]dx=0.084.\label{18}
\end{equation}

\noindent We show the results of the fit for $F^p_2(x) - F^n_2(x)$ and
$F^n_2(x)/F^p_2(x)$ in Figs.~1 and 2 and in Fig.~3 our prediction for
the distribution of the  antiquarks compared to the
data\cit{10}. The behavior of the antiquark distribution is correctly
reproduced for $x>0.1$ but for small $x$ values we don't have the fast
increase shown by the data. This is due to our simplifying assumptions
that the $\tilde x$'s and $\bar x$ are taken to be constant. Actually
when
$x\to 0$, from Pomeron universality, one expects $x\bar u(x) = x\bar
d(x) \not =0$, contrary to the present situation where we took for
quarks and antiquarks a universal $f(x)$ such that $xf(x)$ vanishes when
$x$ goes to zero. Clearly this is not adequate for antiquarks and could
be improved in a more sophisticated version of this approach for example
 by allowing the $\tilde
{x}'$s to depend on $x$. In particular, we underestimate $\bar u$ and
$\bar d$ so we don't satisfy
very accurately the obvious constraints
\begin{equation}
u-\bar u=2\quad\hbox{and}\quad d-\bar d=1.\label{19}
\end{equation}

We now turn to the gluon distribution (see eq.~(\ref{16}) whose only
free
parameter $\tilde x(G)$ has been fixed by the requirement that gluons
carry the fraction of the total momentum not carried by quarks and
antiquarks, that is 0.563, according to eq.~(\ref{18}). We find a very
small value $\tilde
x(G)=-0.012$ and we display in Fig.~4 our prediction which is fairly
consistent with some preliminary indirect experimental determination
{}from direct photon
production\cit{11} and from neutrino deep inelastic scattering\cit{12}
at $Q^2=5$ GeV$^2$.

Going back to quark distributions we can now test our approach by
looking at the predictions we obtain for the polarized proton and
neutron structure functions. So far, all the parameters have been fixed,
except the normalization factor $k$ of $\Delta d$ (see eq.~(\ref{12}))
which was found to be $k=0.787$ after imposing eq.~(\ref{13}). The
results at $Q^2=4$ GeV$^2$ are shown in Fig.~5 for $g^p_1(x)$ and in
Fig.~6 for $g^n_1(x)$. For $g^p_1(x)$ the agreement with the data which
corresponds to $\langle
Q^2\rangle =10$ GeV$^2$, is very satisfactory but no significant $Q^2$
 dependence has been
observed by EMC\cit{5}. For
$g^n_1(x)$ we have the general trend of the data at $\langle Q^2\rangle
=2$ GeV$^2$, which is small,
because there is a strong cancellation between $\Delta u(x)$ weighted
by 1/9 and $\Delta
d(x)$ weighted by 4/9. However we fail to reproduce accurately the
structure around $x=0.1$, in spite of the fact that the driving
contribution $\Delta d(x)$ has a minimum at the correct value because
$\tilde x(u^\downa)=0.143$ (see eq.~(\ref{12})). These predictions are
very encouraging and should certainly be improved.

\section{Quark Parton Model Sum Rules}

There is an intriguing consequence of the hypothesis that the defect in
the Gottfried sum rule follows {\it only} from Pauli principle. In
this framework since in the ``$u^\upa$ bus'', there is less available
space than for the other partons, we expect that the defect in the
Gottfried sum rule arises from the fact that the sea must contribute
less to the $u^\upa$ distribution. This implies from eq.~(\ref{14})

\begin{equation}
\Delta u=\Delta u_{\rm val}+\bar u-\bar d = 2F+\bar u-\bar d =
0.814\pm0.046, \label{20}
\end{equation}

\noindent which is smaller than $\Delta u_{\rm val}$ because $\bar
d>\bar u$. Let us now consider the Bjorken sum rule\cit{16} which reads

\begin{eqnarray}
I_p -I_n & = & \int^1_0 [g^p_1(x)-g^n_1(x)]dx = \frac{1}{6} (\Delta
u + \Delta \bar u -\Delta d- \Delta \bar d) \nonumber \\
& = & \frac{1}{6} (F+D) = \frac{1}{6} (\Delta u_{\rm val} - \Delta
d_{\rm val}) \label{21}
\end{eqnarray}

\noindent where we have used eq.~(\ref{3}) for the last equality. Since
in our approach $\Delta d+\Delta \bar d= \Delta d_{\rm val} =F-D$, in
order to satisfy the Bjorken sum rule we should have $\Delta u+\Delta
\bar u =\Delta u_{\rm val} = 2F$ and therefore $\Delta \bar u$ should be
positive at variance with $\Delta u_{\rm sea}$. Of course, given the
fact that
$\Delta s = \Delta \bar s=0$,\cit{17} this leads for the parton
 Ellis-Jaffe sum
rule\cit{18}

\begin{eqnarray}
I_p &=& \int^1_0g^p_1(x)dx = \frac{2}{9} (\Delta u +\Delta\bar u)
+\frac{1}{18} (\Delta d+\Delta \bar d) \nonumber \\
&=& \frac{F}{2} -\frac{D}{18} = 0.196\pm0.006 \label{22}
\end{eqnarray}

\noindent to be compared with the EMC result $0.126\pm0.015\pm0.009$.
However if one chooses

\begin{equation}\Delta u=2F+\bar u-\bar d, \quad \Delta \bar u=\Delta
u_{\rm sea} = \bar u-\bar d,\quad \Delta d =F-D,\quad \Delta\bar d=0,
\label{23}
\end{equation}

\noindent one gets

\[ I_p=\frac{F}{2} -\frac{D}{18}+\frac{4}{9} (\bar u-\bar d)
=0.134\pm0.015 \]
\noindent and

\begin{equation}
I_n = \frac{1}{3}F -\frac{2}{9} D +\frac{1}{9} (\bar u-\bar d) = -0.024
\pm 0.009 \label{24}
\end{equation}

\noindent in fair agreement with the proton EMC result\cit{5} and with
the neutron SLAC result\cit{13} which is
$I_n=-0.022\pm0.011$.\footnote{A much less accurate CERN
experiment\cit{19} leads to a different result, i.e.
$I_n=-0.08\pm0.04\pm0.04$.}

This choice (eq.~(\ref{23})) which is obtained by assuming the crucial
role of Pauli principle for parton distributions implies that the
theoretical framework for the derivation of the Bjorken sum rule should
be reconsidered.

\section {Concluding remarks}

We have obtained a good description of the main features of unpolarized
and polarized deep inelastic scattering by using Fermi-Dirac
distributions for quark and antiquark partons. This supports the idea,
previously proposed\cit{7}, that Pauli exclusion principle plays a
crucial role in deep inelastic phenomena and leads to the intringuing
conclusion that partons interact incoherently but are bound to obey
Fermi-Dirac statistics in the $x$ variable. Concerning the parameters
of these distributions, we have found a universal temperature $\bar x$
for quarks, antiquarks and gluons, whereas the thermodynamical
potentials $\tilde x(p)$ are positive for quarks, negative for
antiquarks and small for gluons. As expected, we find that $\tilde
x(u^{\upa})$ dominates over all the other ones, so the "$u^{\upa}$ bus"
is rather full which is naturaly related to the defect of some quark
parton model sum rules. In its present form, our approach fails to
reproduce the fast increase of the antiquark distribution at very low
$x$ and this deserves some further comments. One obvious reason is that
within the parton model approximation the distributions are depending
on $x$ only, not on $Q^2$, and $\bar x$ and $\tilde x(p)$ were taken to
be constants. Instead, we could have assumed that they depend also on
the final hadronic invariant mass $(P+q)^2=M^2+Q^2((1-x)/x)$ which
varies rapidly at small $x$. We may observe that this fast increase of
the distributions at small $x$ was advocated in the framework of the
multiperipheral approach of deep inelastic scattering, where a singular
behaviour $x^{-3/2}$ was suggested\cit{20} and this trend was recently
observed in the data\cit{21} for the first time. In this case one can
speak pictorially of a liquid of partons for which the Fermi-Dirac gas
description is inadequate. The parton distributions contain two phases,
a gas contributing to the non singlet part which dominates at moderate
and large $x$ and a liquid contributing to the singlet part which
prevails at low $x$. This is not a new situation and it corresponds to
the well known two components picture, with ordinary Regge trajectories
related to resonances and the Pomeron. Finally, we would like to add
that we have been recently aware of a work\cit{22} where deep inelastic
structure functions are calculated in a statistical model of the
nucleon considered as a gas of quarks and gluons in the framework of
the MIT bag model. Despite the commun reference to quantum statistics,
the two approaches are rather different and in ref.[22] only
unpolarized distributions have been considered.

\section*{Acknowledgments}

One of us (JS) is gratefull for kind hospitality at the High Energy
Theory Group, Brookhaven National Laboratory, where part of this work
was done. We also thank E. Hughes, S.R. Mishra, M. Shaevitz and R.
Windmolders for providing us with numerical values on data from E142,
CCFR and NMC Collaborations.

\newpage

\section*{Figure Captions}

\begin{itemize}
\item[Fig.1] The difference $F^p_2(x)-F^n_2(x)$ at $Q^2=4$ GeV$^2$
versus $x$. Data are from ref.[9] and the solid line is the result of
our fit.

\item[Fig.2] The ratio $F^n_2(x)/F^p_2(x)$ at $Q^2=4$ GeV$^2$ versus
$x$. Data are from ref.[9] and the solid line is the result of our fit.

\item[Fig.3] The antiquark contribution $x\bar q(x)=x\bar u(x)+ x\bar
d(x)+x\bar s(x)$ at $Q^2=3$ GeV$^2$ (open circles) and $Q^2=5$ GeV$^2$
(full triangles) versus $x$. Data are from ref.[10] and solid line is
the result of our fit.

\item[Fig.4] The gluon distribution $xG(x)$ at $Q^2=5$ GeV$^2$ versus
$x$. Data are from ref.[11] (area between dashed lines) and ref.[12]
(area between small dashed lines) and solid line is the result of our
calculation at $Q^2=4$ GeV$^2$.

\item[Fig.5] $xg^p_1(x)$ at $<Q^2>=10$ GeV$^2$ versus $x$. Data are
{}from ref.[5] and solid line is our prediction at $Q^2=4$ GeV$^2$.

\item[Fig.6] $xg^n_1(x)$ at $<Q^2>=2$ GeV$^2$ versus $x$. Data are from
ref.[13] and solid line is our prediction at $Q^2=4$ GeV$^2$.
\end{itemize}

\end{document}